\documentclass[aps,preprint,showpacs,amsmath,amssymb]{revtex4-1}
\usepackage{graphicx}
\usepackage{dcolumn}
\usepackage{bm}
\usepackage{hyperref}
\begin{document}

\title{Sudden change in dynamics of genuine multipartite entanglement of cavity-reservoir qubits}
\author{Mazhar Ali$^{1,2}$, and A. R. P. Rau$^3$}
\affiliation{$^1$Naturwissenschaftlich-Technische Fakult\"at, Universit\"{a}t Siegen, Walter-Flex-Stra\ss e 3, 57068 Siegen, Germany\\
$^2$Department of Electrical Engineering, COMSATS Institute of Information Technology, 22060 Abbottabad, Pakistan\\
$^3$Department of Physics and Astronomy, Louisiana State University, Baton Rouge, 70803 Louisiana, USA}


\begin{abstract}
We study the dynamics of genuine multipartite entanglement for a system of four qubits. 
Using a computable entanglement monotone for multipartite systems, we investigate the as yet
unexplored aspects of a cavity-reservoir system of qubits. For one specific initial 
state, we observe a sudden transition in the dynamics of genuine entanglement for the four qubits. This sudden 
change occurs only during a time window where neither cavity-cavity qubits nor reservoir-reservoir 
qubits are entangled. We show that this sudden change in dynamics of this specific state 
is extremely sensitive to white noise.  
\end{abstract}


\maketitle

\section{Introduction}\label{S-intro}

Multipartite entanglement is one of the peculiar features of quantum physics and 
it is important to study its characterization and its dynamics under decoherence \cite{Horodecki-RMP-2009, gtreview}.
Several research efforts have been undertaken to study the dynamics of bipartite and multipartite entanglement under 
decoherence \cite{YE-work, lifetime, acindec, bipartitedec,lowerbounds, Lastra-PRA75-2007, Guehne-PRA78-2008, furtherdec, Ali-JPB-2014}.
Previously, with the exception of Ref. \cite{Ali-JPB-2014}, only bipartite aspects of the entanglement of 
several parties have been considered \cite{bipartitedec}. However, this can 
only give a partial characterization, since multipartite entanglement is known to be different from entanglement 
between all bipartitions \cite{gtreview}. Also, the theory of multipartite 
entanglement is still not fully developed, so that for many cases one can only set lower bounds on entanglement, but not obtain the value itself \cite{lowerbounds}. The exact calculation of a multipartite entanglement measure has been done only for special states and decoherence models \cite{Guehne-PRA78-2008}.

In this paper, we study entangled cavity photons interacting with independent reservoirs. The presence or absence of a photon in a cavity 
defines one qubit, whereas no photon or normalized collective state with single excitation in the reservoir defines the second qubit. 
Two such entangled cavities interacting with two independent reservoirs effectively forms a system of four qubits. Some interesting aspects of this system have been investigated before \cite{Lopez-PRL101-2008}. However, due to the unavailability of a genuine entanglement monotone, 
only bipartite aspects were explored. One important feature of this study was the demonstration of the relations between disappearance and appearance of two-qubit entanglement from cavities to reservoirs. In particular, it was shown that if 
entanglement between the cavities decays asymptotically, then the corresponding entanglement between reservoirs starts to grow 
immediately after the interaction starts. But, for the case when entanglement between cavities disappears in finite time, the entanglement between reservoirs does not grow immediately after the interaction starts but it may appear either before, simultaneously, or 
even after the disappearance of entanglement between cavities. For the last possibility, there is a time window where neither cavities nor 
reservoirs are entangled among themselves. We find that this time window can have a profound effect on the dynamics of genuine multipartite entanglement.

For one specific initial state, and only during this time window, genuine entanglement exhibits a sudden 
change in its dynamics and freezes onto a constant value in this time window. But, we do not find such behavior for 
other initial states. Therefore, this sudden change in dynamics of genuine entanglement is not a general feature. 
Our study is enabled by recent progress in the theory of multipartite 
entanglement, especially the computable entanglement monotone for genuine multipartite entanglement \cite{Bastian-PRL106-2011}. 
Interestingly, similar behavior has also been reported recently for other quantum correlations like quantum discord 
\cite{Maziero-PRA80-2009, Mazzola-PRL104-2010, Mazzola-IJQI9-2011, Pinto-PRA88-2013}.

This paper is organized as follows. In section \ref{Sec:Model}, we briefly describe our model of interaction. We review the concept of 
genuine entanglement and multipartite negativity in section \ref{Sec:GME}. We present our main results in section \ref{Sec:results}. 
Finally, we conclude the work in section \ref{Sec:conc}. 

\section{Open-system dynamics of two qubits coupled to statistically independent reservoirs} \label{Sec:Model}

In this section, we illustrate our model and the basic equations of motion governing our system of interest. We consider our qubits as 
two uncoupled cavity modes with upto one photon. Each mode interacts independently with its own reservoir. 
The Hamiltonian describing the interaction between a single cavity mode and a $N$-mode reservoir is given by \cite{Lopez-PRL101-2008}
\begin{eqnarray}
\hat{H} = \hbar \omega \hat{a}^\dagger \hat{a} + \hbar \sum_{k = 1}^N \omega_k \hat{b}^\dagger \hat{b} 
+ \hbar \sum_{k = 1}^N g_k \big( \hat{a} \hat{b}_k^\dagger + \hat{b}_k \hat{a}^\dagger \big).
\label{Eq:Hamil}
\end{eqnarray}
The first term describes the single cavity mode, the second term is the $N$-mode reservoir, and the third term describes the interaction between cavity 
and reservoir. As we are interested in the situation where a cavity mode contains only a single photon and its corresponding reservoir 
is in the vacuum mode, the combined state before the interaction can be written as
\begin{eqnarray}
|\psi(0)\rangle_{CR} = |1 \rangle_C \otimes |\bar{{\bf 0}}\rangle_R, \label{Eq:IS}  
\end{eqnarray}
where $ |\bar{{\bf 0}}\rangle_R = \Pi_{k=1}^N |{\bf 0}_k\rangle_R$ is the collective vacuum state of $N$-modes of reservoir $R$. 
The time evolution of this state using Hamiltonian (\ref{Eq:Hamil}) leads to 
\begin{eqnarray}
|\psi(t)\rangle_{CR} = \xi(t) |1\rangle_C \, |\bar{{\bf 0}}\rangle_R + \sum_{k=1}^N \lambda_k(t) \, |0\rangle_C \, |{\bf 1}_k\rangle_R \, ,
\label{Eq:TE}
\end{eqnarray}
where the reservoir state $|{\bf 1}_k\rangle_R$ describes the presence of a single photon in mode $k$. The probability amplitude 
$\xi(t)$ converges to $\xi(t) = e^{- \kappa t/2}$ in the limit of large $N$, i.e., $N \to \infty$. Eq.(\ref{Eq:TE}) can be written as
\begin{eqnarray}
|\psi(t)\rangle_{CR} = \xi(t) |1\rangle_C \, |\bar{{\bf 0}}\rangle_R + \chi(t) |0\rangle_C \, |\bar{{\bf 1}} \rangle_R \, ,
\label{Eq:TEES}
\end{eqnarray}
where $|\bar{{\bf 1}} \rangle_R = (1/\chi(t)) \sum_{k=1}^N \lambda_k(t) |{\bf 1}_k\rangle$, and the probability amplitude $\chi(t)$ 
converges to $\chi(t) = \sqrt{1-e^{- \kappa t}}$ for large $N$. Eq.(\ref{Eq:TEES}) describes an effective two-qubit 
system \cite{Lopez-PRL101-2008}.

After defining the basic equation of motion, we can now study the joint time evolution of two qubits with their corresponding reservoirs assumed 
initially to be in the vacuum state. Taking two qubits in an arbitrary X-state \cite{YE-work, ARPR09}, we write 
\begin{eqnarray}
\rho_{tot} (0) = \rho_X \otimes |\bar{0}_{r_1} \bar{0}_{r_2}\rangle\langle\bar{0}_{r_1} \bar{0}_{r_2}| \, , 
\label{Eq:XIS}
\end{eqnarray}
where the density matrix of a general two-qubit X-state is 
\begin{eqnarray}
\rho_X = \left( 
\begin{array}{cccc}
\rho_{11} & 0 & 0 & \rho_{14} \\ 
0 & \rho_{22} & \rho_{23} & 0 \\ 
0 & \rho_{32} & \rho_{33} & 0 \\
\rho_{41} & 0 & 0 & \rho_{44}
\end{array}
\right).
\label{Eq:Xs}
\end{eqnarray}
Eq.~(\ref{Eq:Xs}) describes a quantum state provided the unit trace and positivity conditions $\sum_{i=1}^4 \rho_{ii} = 1$,  
$ \rho_{22} \rho_{33} \geq |\rho_{23}|^2$, and $ \rho_{11} \rho_{44} \geq |\rho_{14}|^2$ are fulfilled. X-states are entangled if and only 
if either $\rho_{22} \rho_{33} < |\rho_{14}|^2$ or $\rho_{11} \rho_{44} < |\rho_{23}|^2$. 
The orthonormal photonic eigenstates 
$|1\rangle = |0 \rangle_A \otimes |0\rangle_B$,
$|2\rangle = |0\rangle_A \otimes |1\rangle_B$,
$|3\rangle = |1\rangle_A \otimes |0\rangle_B$,
$|4\rangle = |1\rangle_A \otimes |1\rangle_B$ 
form the (computational) basis of the four dimensional Hilbert space of the two qubits. The time evolution of Eq.~(\ref{Eq:XIS}), 
according to Eq.~(\ref{Eq:TEES}) can be determined easily. After partially tracing over the reservoirs degrees of freedom, the time evolved 
density matrix for system qubits becomes
\begin{widetext}
\begin{eqnarray}
\rho_{C_1 C_2}(t) = \left( 
\begin{array}{cccc}
\rho_{11}(t) & 0 & 0 & \rho_{14} \, \xi^2(t) \\ 
0 & (\rho_{22} + \rho_{44} \, \chi^2(t) ) \, \xi^2(t) & \rho_{23} \, \xi^2(t) & 0 \\ 
0 & \rho_{32} \, \xi^2(t) & (\rho_{33} + \rho_{44} \, \chi^2(t)) \, \xi^2(t) & 0 \\
\rho_{41} \, \xi^2(t) & 0 & 0 & \rho_{44} \, \xi^4(t)
\end{array}
\right),
\label{Eq:TECXs}
\end{eqnarray}
\end{widetext}
where $\rho_{11}(t) = \rho_{11} + ( \rho_{22} + \rho_{33} + \rho_{44} \, \chi^2(t) ) \, \chi^2(t)$. Similarly, after tracing over cavity 
qubits, we can write the density matrix for environment qubits as
\begin{widetext}
\begin{eqnarray}
\sigma_{R_1 R_2}(t) = \left( 
\begin{array}{cccc}
\sigma_{11}(t) & 0 & 0 & \rho_{14} \, \chi^2(t) \\ 
0 & (\rho_{22} + \rho_{44} \, \xi^2(t) ) \, \chi^2(t) & \rho_{23} \, \chi^2(t) & 0 \\ 
0 & \rho_{32} \, \chi^2(t) & (\rho_{33} + \rho_{44} \, \xi^2(t)) \, \chi^2(t) & 0 \\
\rho_{41} \, \chi^2(t) & 0 & 0 & \rho_{44} \, \chi^4(t)
\end{array}
\right),
\label{Eq:TERXs}
\end{eqnarray}
\end{widetext}
where $\sigma_{11}(t) = \rho_{11} + ( \rho_{22} + \rho_{33} + \rho_{44} \, \xi^2(t) ) \, \xi^2(t)$.

\section{Genuine multipartite entanglement and multipartite negativity} 
\label{Sec:GME}

In this section, we review the basic definitions for genuine multipartite 
entanglement and multipartite negativity that are currently available in the literature. We discuss the main ideas by considering three 
parties $A$, $B$, and $C$, generalization to more parties being straightforward. A state is called separable with respect 
to some bipartition, say, $A|BC$, if it is a mixture of product states with respect 
to this partition, that is, 
$\rho = \sum_j \, p_j \, |\psi_A^j \rangle\langle \psi_A^j| \otimes |\psi_{BC}^j \rangle\langle \psi_{BC}^j|$, 
where the $p_j$ form a probability distribution. We denote these states as $\rho_{A|BC}^{sep}$. 
Similarly, we can define separable states for the two other bipartitions, $\rho_{B|CA}^{sep}$ and $\rho_{C|AB}^{sep}$. 
Then a state is called biseparable if it can be written as a mixture of states which are separable with respect 
to different bipartitions, that is 
\begin{eqnarray}
 \rho^{bs} = \tilde{p}_1 \, \rho_{A|BC}^{sep} + \tilde{p}_2 \, \rho_{B|AC}^{sep} + \tilde{p}_3 \, \rho_{C|AB}^{sep}\,,
\end{eqnarray}
with $\tilde{p}_1 +\tilde{p}_2 +\tilde{p}_3 = 1$.
Finally, a state is called genuinely multipartite entangled if it is not biseparable. In the rest of this paper, we always mean genuine
multipartite entanglement when we talk about multipartite entanglement. 

Recently, a powerful technique has been advanced to detect and characterize multipartite entanglement \cite{Bastian-PRL106-2011}. 
The technique is based on using positive partial transpose mixtures (PPT mixtures). We recall that a two-party state 
$\rho = \sum_{ijkl} \, \rho_{ij,kl} \, |i\rangle\langle j| \otimes |k\rangle\langle l|$ is PPT if its partially transposed matrix 
$\rho^{T_A} = \sum_{ijkl} \, \rho_{ji,kl} \, |i\rangle\langle j| \otimes |k\rangle\langle l|$ has no negative eigenvalues. 
It is known that separable states are always PPT \cite{peresppt}. The set of separable states with respect to some partition 
is therefore contained in a larger set of states which has a positive partial transpose for that bipartition. 

We denote the states which are PPT with respect to fixed bipartition by $\rho_{A|BC}^{PPT}$, $\rho_{B|CA}^{PPT}$, 
and $\rho_{C|AB}^{PPT}$ and ask whether a state can be written as
\begin{eqnarray}
\rho^{PPTmix} = q_1 \, \rho_{A|BC}^{PPT} + q_2 \, \rho_{B|AC}^{PPT} + q_3 \, \rho_{C|AB}^{PPT}\,.
\end{eqnarray}
The mixing of PPT states is called a PPT mixture. As any biseparable state is a PPT mixture, therefore any state which is not 
a PPT mixture is guaranteed to be genuinely multipartite entangled. The main advantage of considering PPT mixtures instead 
of biseparable states comes from the fact that PPT mixtures can be fully characterized by the method of semidefinite programming 
(SDP), a standard  method in convex optimization \cite{sdp}. Generally the set of PPT mixtures is a very good approximation to the 
set of biseparable states and delivers the best known separability criteria for many cases; however, there are multipartite entangled 
states which are PPT mixtures \cite{Bastian-PRL106-2011}.

We briefly describe SDP. It was shown \cite{Bastian-PRL106-2011} that a state is a PPT mixture iff the following optimization problem 
\begin{eqnarray}
\min {\rm Tr} (\mathcal{W} \rho)
\end{eqnarray}
with constraints that for all bipartition $M|\bar{M}$
\begin{eqnarray}
\mathcal{W} = P_M + Q_M^{T_M},
 \quad \mbox{ with }
 0 \leq P_M\,\leq 1 \mbox{ and }
 0 \leq  Q_M  \leq 1\, 
\end{eqnarray}
has a positive solution. The constraints just state that the considered operator $\mathcal{W}$ is a decomposable entanglement
witness for any bipartition. If this minimum is negative, then $\rho$ is not a PPT mixture and hence is genuinely multipartite 
entangled. Since this is a semidefinite program, the minimum can be efficiently computed and the optimality of the solution can
be certified \cite{sdp}. For solving the SDP we used the programs YALMIP and SDPT3 \cite{yalmip}, a ready-to-use implementation 
being freely available \cite{pptmix}.

It is important that this approach can be used to quantify genuine entanglement. In fact, the absolute value of the above 
minimization was shown to be an entanglement monotone for genuine multipartite entanglement \cite{Bastian-PRL106-2011}. 
We denote this measure by $E(\rho)$ or $E$-monotone in this paper. For bipartite systems, this monotone is equivalent to 
{\it negativity} \cite{Vidal-PRA65-2002}. 
For a system of qubits, this measure is bounded by  $E(\rho) \leq 1/2$ \cite{bastiangraph}.
%

\section{Results} \label{Sec:results}

In this section, we present our main results for various initial states of two cavity qubits. 

$(1)$ First let us consider the pure state
\begin{eqnarray}
|\psi(0)\rangle =  \, \alpha \, |0 \, 0 \rangle \, + \, \beta \, |1 \, 1\rangle \, . 
\label{Eq:JIS}
\end{eqnarray}
It is known that the time-evolved density matrix for cavity qubits looses its entanglement (``entanglement sudden death") in a finite time $t_{ESD}$ for 
$\alpha < \beta$ \cite{YE-work}. The consequences on the entanglement transferred to reservoir qubits are quite 
interesting. The entanglement between reservoirs may appear at a time $t_{ESB}$ (``entanglement sudden birth") depending on the relation between 
$\alpha$ and $\beta$. The times for loss of entanglement from cavities and appearance in reservoirs are given 
as \cite{Lopez-PRL101-2008}
\begin{eqnarray}
t_{ESD} &=& \frac{1}{\kappa} \ln \bigg(\frac{\beta}{\beta -\alpha}\bigg) \nonumber \\
t_{ESB} &=& \frac{1}{\kappa} \ln \bigg(\frac{\beta}{\alpha} \bigg) \, .
\label{Eq:TECBR}
\end{eqnarray}
From these relations, it is clear that these times are equal for $\beta = 2 \, \alpha$. The reservoirs may get entangled before the 
cavities disentangle for $\beta < 2 \, \alpha$ or the reservoirs' entanglement may come after the cavities have disentangled 
for $\beta > 2 \, \alpha$. This last possibility has the peculiarity of a time window where both cavity qubits and reservoir 
qubits are disentangled. We will explore this region later. 

\begin{figure}[h]
\centering
\scalebox{2.25}{\includegraphics[width=1.95in]{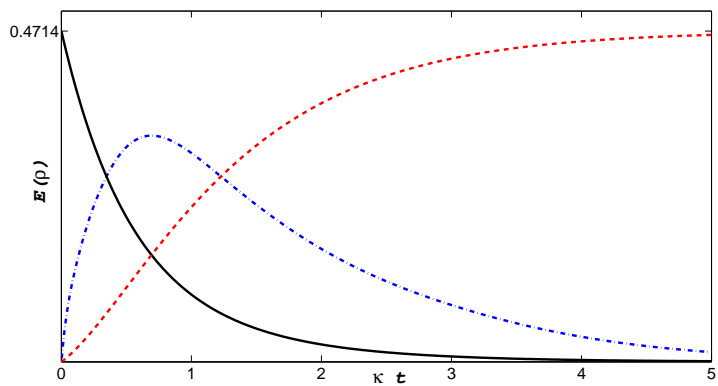}}
\caption{$E$-monotone is plotted against parameter $\kappa t$ for cavity-cavity qubits (black solid line), reservoir-reservoir 
qubits (red dashed line) and all four qubits (blue dashed-dotted line) for $\alpha = \sqrt{2/3}$ and $\beta = \sqrt{1/3}$. See text for more description.}
\label{FIG:CR1}
\end{figure}
First we examine the situation that leads to an asymptotic decay of 
cavity qubits. Figure (\ref{FIG:CR1}) shows the $E$-monotone plotted against parameter $\kappa t$ with 
initial condition $\alpha = \sqrt{2/3}$ and $\beta = \sqrt{1/3}$. 
The solid line is for cavity-cavity qubits, whose entanglement decays asymptotically. The dashed line is the entanglement of 
reservoir-reservoir qubits. It is clear that for such an initial density matrix when there is no finite time disentanglement in cavity qubits, there is also no sudden 
appearance of entanglement between the reservoirs. The dashed-dotted line denotes the $E$-monotone for multipartite entanglement of four qubits. 
We find that all four qubits get multipartite entangled immediately after the interaction starts and remain so until 
infinity where all entanglement is transferred to the reservoirs alone. The initial value of the $E$-monotone for the cavities exactly equals the final value for the reservoirs, and that for all four qubits rises to a maximum and decays asymptotically.

\begin{figure}[h]
\centering
\scalebox{2.25}{\includegraphics[width=1.95in]{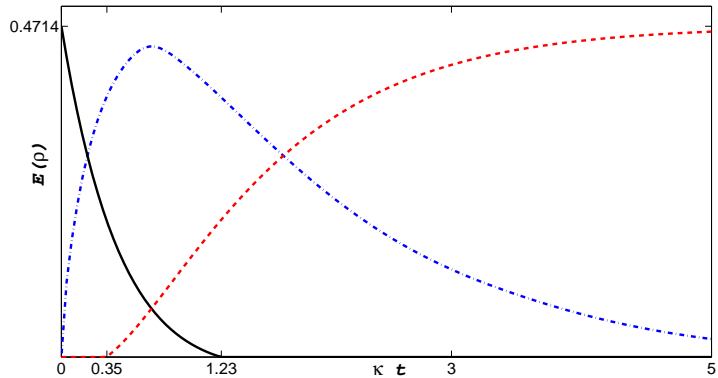}}
\caption{$E$-monotone is plotted against parameter $\kappa t$ for cavity-cavity qubits (black solid line), reservoir-reservoir 
qubits (red dashed line) and all four qubits (blue dashed-dotted line) for $\alpha = \sqrt{1/3}$ and $\beta = \sqrt{2/3}$.}
\label{FIG:CR2}
\end{figure}
We turn next to the case when there is finite time disentanglement in the cavity qubits and a corresponding sudden appearance of 
entanglement between the reservoir qubits. By starting with initial condition $\alpha = \sqrt{1/3}$ and $\beta = \sqrt{2/3}$, we expect that 
reservoir-reservoir entanglement will appear before there is complete disentanglement in cavity qubits since $\beta < 2 \, \alpha$. This is indeed 
what we observe in Figure (\ref{FIG:CR2}) for the $E$-monotone. The solid line denotes entanglement 
of cavity-cavity qubits and comes to an end at $\kappa t \approx 1.23$, whereas entanglement of reservoirs (dashed line) appears at 
about $\kappa t \approx 0.35$. The dashed-dotted line is for four-qubits entanglement. We observe that all four qubits get multipartite 
entangled immediately after the interaction starts and they remain entangled for long time. 

\begin{figure}[h]
\centering
\scalebox{2.25}{\includegraphics[width=1.95in]{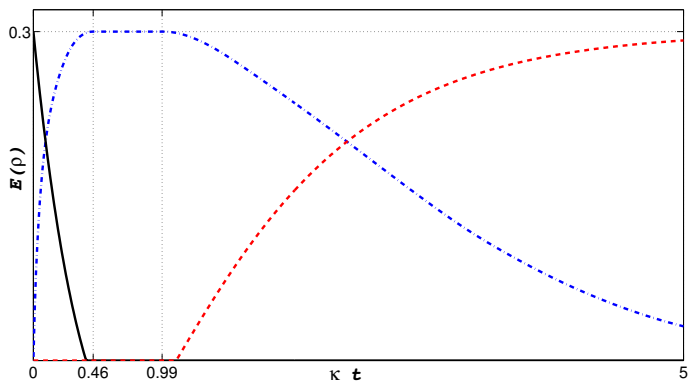}}
\caption{$E$-monotone is plotted against parameter $\kappa t$ for cavity-cavity qubits (black solid line), reservoir-reservoir 
qubits (red dashed line) and all four qubits (blue dashed-dotted line). In the time window when there is no entanglement between cavity 
qubits and reservoir qubits, the multipartite entanglement exhibits an abrupt change to a value that remains constant in that window.}
\label{FIG:CR3}
\end{figure}
We now focus on the peculiar situation with initial condition $\beta > 2 \ \alpha$. As noted before, this choice of parameters give rise 
to a time window where neither cavity qubits nor reservoir qubits are entangled. We first start with $\alpha = \sqrt{1/10}$ and 
$\beta = 3 \, \sqrt{1/10}$. Figure (\ref{FIG:CR3}) depicts the $E$-monotone for cavity-cavity qubits (solid line), reservoir-reservoir
qubits (dashed line), and four-qubit entanglement (dashed-dotted line). 
The entanglement of cavity qubits (solid line) comes to an end at $\kappa t \approx 0.41$, whereas entanglement in the reservoirs (dashed line) 
appear at $\kappa t \approx 1.1$. Hence during the time window, that is, between 
$ \kappa t \approx 0.41$ and $\kappa t \approx 1.1$, neither the cavity qubits nor the reservoir qubits are entangled. On the other hand, 
the genuine entanglement (dashed-dotted line) reaches a maximum value $E(\rho_{CCRR}) = 0.3$ (which is equal to amount of monotone of cavity qubits 
at $\kappa t = 0$, that is, $E(\rho_{CC}(0)) = 0.3$) and exhibits a sudden change in its dynamics shortly after the point where the cavity qubits are 
disentangled, at $\kappa t \approx 0.46$. The genuine entanglement maintains its constant value of $0.3$ shortly before a point where entanglement suddenly 
appears in the reservoirs at $\kappa t \approx 0.99$. After this point, it starts decaying asymptotically.

\begin{figure}[h]
\centering
\scalebox{2.25}{\includegraphics[width=1.95in]{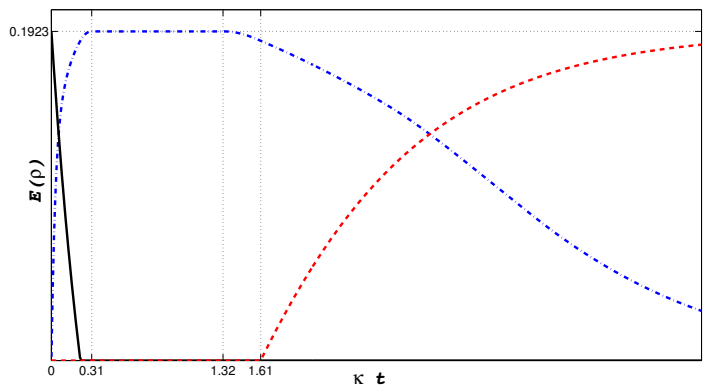}}
\caption{Same caption as Figure (\ref{FIG:CR3}). See the text for more descriptions.}
\label{FIG:CR4}
\end{figure}
We take another example with a broader time window, that is, with initial condition $\alpha = \sqrt{1/26}$ and 
$\beta = 5 \, \sqrt{1/26}$. Figure (\ref{FIG:CR4}) depicts the $E$-monotone for three sets of density matrices. It is obvious from this 
figure that cavity-cavity entanglement (solid line) comes to an end at $\kappa t \approx 0.23$. 
In the range $0.31 \leq \kappa t \leq 1.32$, we again see the peculiar dynamics of multipartite entanglement 
denoted by dashed-dotted line. The multipartite negativity starts growing immediately after the interaction starts and at 
$\kappa t \approx 0.31$ it exhibits an abrupt change change. It attains its maximum value of $E \approx 0.1923$ (equal to initial value of monotone  
for cavity qubits) and maintains it till $\kappa t \approx 1.32$ after that genuine entanglement starts decaying asymptotically. 
The reservoirs get entangled at $\kappa t \approx 1.61$. 

$(2)$ We take the mixed state density matrix of Werner states, which are an important single parameter class of states, given as 
\begin{eqnarray}
\rho_p = p \, |\Phi \rangle\langle\Phi| + \frac{(1-p)}{4} \mathbb{I}_4 \, , \label{Eq:8}
\end{eqnarray}
where $p \in [0,1]$ and $|\Phi\rangle = 1/\sqrt{2} (|0,0\rangle + |1,1\rangle)$ is the maximally entangled pure 
Bell state. It is well known that Werner states are entangled for $p \in (1/3,1]$ and separable for $p \leq 1/3$. The time of disentanglement 
for the Werner state (\ref{Eq:8}) is given by
\begin{eqnarray}
t_{ESD} (\rho_p) = \frac{1}{\kappa} \ln\bigg[\frac{1 + p}{2(1-p)}\bigg]\, . 
\label{Eq:9}
\end{eqnarray}
We note that this time is always finite for $p \in (1/3,1)$, however for $ p = 1$, there is no abrupt disentanglement and maximally entangled Bell state 
loses its entanglement at infinity. The time of sudden birth of entanglement among reservoir qubits is given by
\begin{eqnarray}
t_{ESB} (\rho_p) = \frac{1}{\kappa} \ln\bigg[\frac{1 + p}{-1 + 3 p} \bigg]\, . 
\label{Eq:10}
\end{eqnarray}
In the range $ p \in (1/3,1)$, this time is finite and becomes zero for $p = 1$, which simply means that when there is no finite time 
disentanglement among cavity qubits, there is also no sudden birth of entanglement among reservoirs and entanglement appears among reservoir 
qubits immediately after the interaction starts. It is not difficult to check that for $ p = 3/5$, both these times are equal and sudden death and sudden 
birth occur simultaneously. For $p > 3/5$, sudden birth comes earlier than sudden death, and for $p < 3/5$ entanglement among reservoirs appear after there is no 
entanglement between cavity qubits. 

\begin{figure}[h]
\scalebox{2.20}{\includegraphics[width=1.95in]{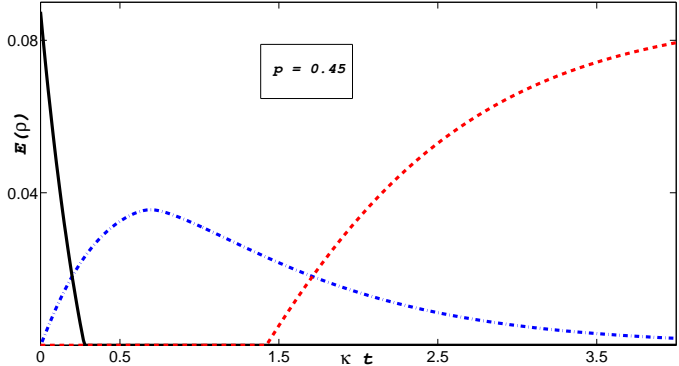}}
\caption{Entanglement monotone for cavity qubits (solid line), reservoir qubits (dashed line), and collective state of four qubits 
(dashed-dotted line) is plotted against parameter $\kappa t$ for Werner states with $p = 0.45$.}
\label{FIG:WSc1}
\end{figure}
To check whether sudden change in dynamics of genuine multipartite entanglement occurs for these states, 
we take $p = 0.45$. In Figure \ref{FIG:WSc1}, we plot the entanglement monotone for cavity-cavity qubits (black solid line), reservoir-reservoir 
qubits (red dashed line), and combined state of four qubits (blue dashed-dotted line). As expected, we find that cavity qubits lose their 
entanglement at $\kappa t \approx 0.28$ whereas entanglement among reservoirs appear at $\kappa t \approx 1.44$, creating a time window 
between two events where neither cavity qubits not reservoir qubits are entangled. However, we do not observe any sudden change in dynamics of 
genuine entanglement of four qubits (blue dashed-dotted line). The phenomenon of an abrupt transition to a constant value seems, therefore, not to be a generic feature but dependent on the initial state. 

\begin{figure}[h]
\scalebox{2.20}{\includegraphics[width=1.95in]{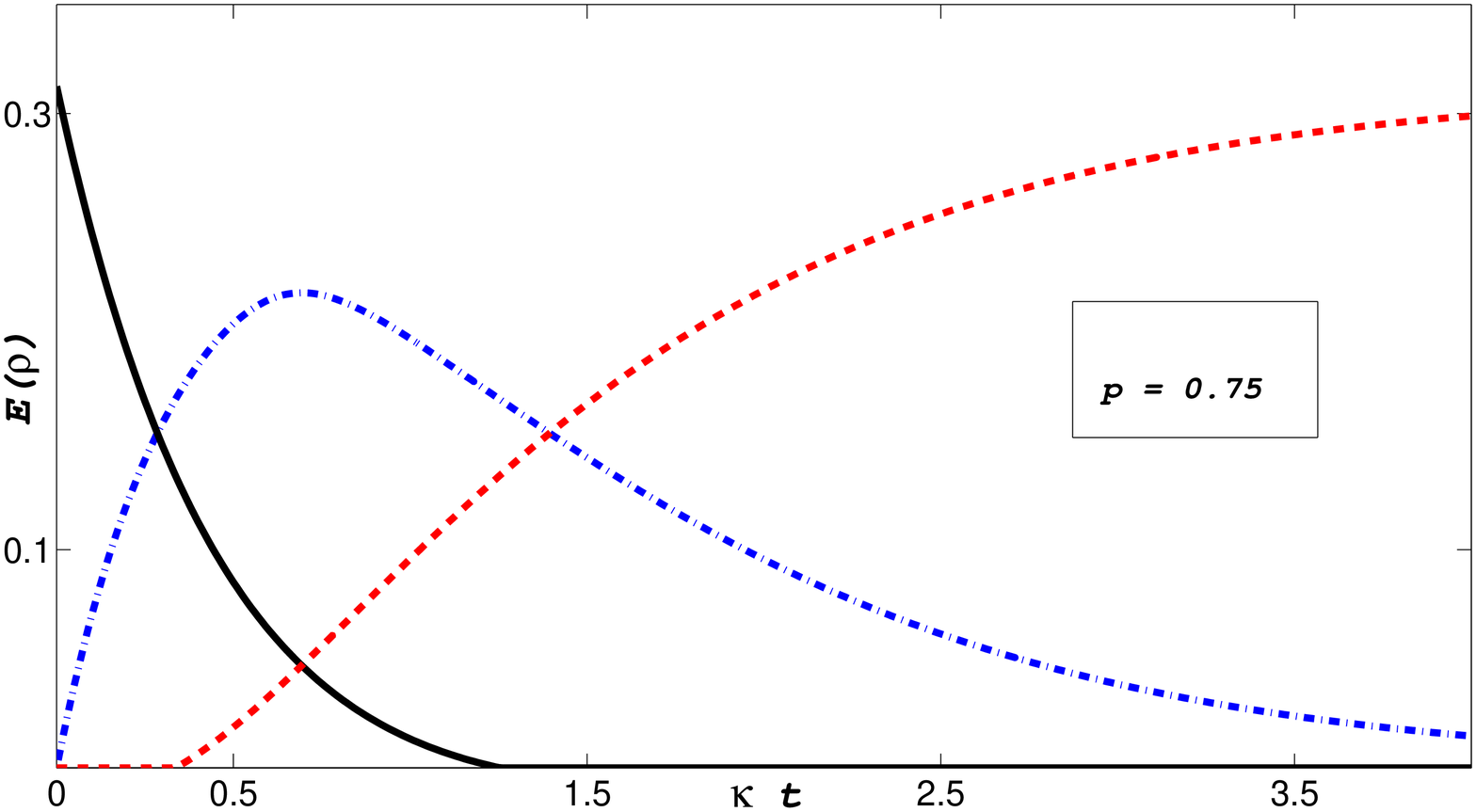}}
\caption{Same caption as Figure (\ref{FIG:WSc1}) but for $p = 0.75$.}
\label{FIG:WSc2}
\end{figure}

Figure \ref{FIG:WSc2} shows a similar study with $p = 0.75$. Entanglement among cavity qubits is lost at $\kappa t \approx 1.26$, 
whereas it appears among reservoirs at $\kappa t \approx 0.34$. The four-qubit entanglement gradually increases to a maximum value and then 
starts decaying asymptotically.

$(3)$ As another example, we consider an initial entangled state of the form 
\begin{eqnarray}
\rho_a = \frac{1}{3} ( a |1,1\rangle\langle1,1| + 2 |\Psi\rangle\langle\Psi| + (1-a) |0,0\rangle\langle0,0|)\,,
\label{Eq:YES}
\end{eqnarray}
with $|\Psi\rangle = (|0,1\rangle + |1,0\rangle)/\sqrt{2}$ and $0 \leq a \leq 1$. We find that this state exhibits finite-time 
disentanglement for $a > 1/3$, whereas for other range of values, there is no sudden death and also no sudden birth of entanglement. 
The time of sudden death for the state (\ref{Eq:YES}) is given by
\begin{eqnarray}
t_{ESD} (\rho_a) = \frac{1}{\kappa} \ln\bigg[\frac{a + a^2 + \sqrt{2 a^2 - a^3 + a^4}}{-1 + 3 a} \, \bigg]\, . 
\label{Eq:TESDa}
\end{eqnarray}
We note that this time is always finite for $a > 1/3$. Whereas the corresponding time of sudden birth is given by
\begin{eqnarray}
t_{ESB} (\rho_a) = \frac{1}{\kappa} \ln\bigg[\frac{a + \sqrt{2 a^2 - a^3 + a^4}}{1 - a + a^2} \, \bigg]\, . 
\label{Eq:TESBa}
\end{eqnarray}
It is not difficult to find that this time is zero for $a \leq 1/3$. It turns out that for $a = 2/3$, both of these times are equal and sudden death and 
sudden birth occur simultaneously. For $a > 2/3$, sudden death comes earlier than sudden birth and vice versa.

\begin{figure}[h]
\scalebox{2.20}{\includegraphics[width=1.95in]{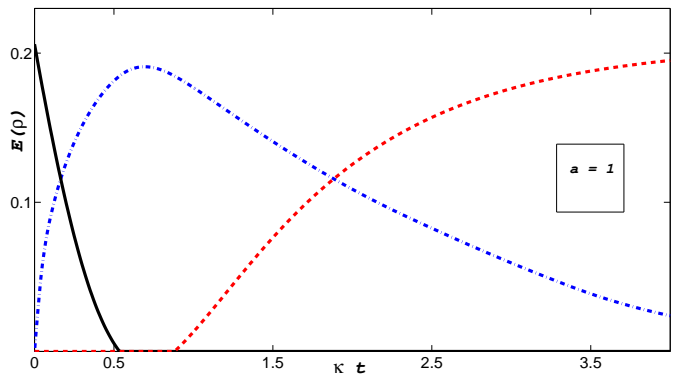}}
\caption{Entanglement monotone is plotted for cavity-cavity (solid line), reservoir-reservoir (dashed line), and four qubits  
(dashed-dottd line) for states (\ref{Eq:YES}) for $a = 1$.}
\label{FIG:YEc1}
\end{figure}
In Figure \ref{FIG:YEc1}, we plot the entanglement monotones for the initial states (\ref{Eq:YES}) with $a = 1$. The cavity qubits lose their entanglement at 
$\kappa t \approx 0.535$, whereas the reservoirs get entangled at $\kappa t \approx 0.8814$, again showing the time window with no entanglement for 
cavity and reservoir qubits. One again, the dynamics of genuine entanglement for four qubits show no sudden change and decays asymptotically 
after reaching its maximum value. This gives another example to indicate that the sudden change is not a generic feature. 

\begin{figure}[h]
\scalebox{2.20}{\includegraphics[width=1.95in]{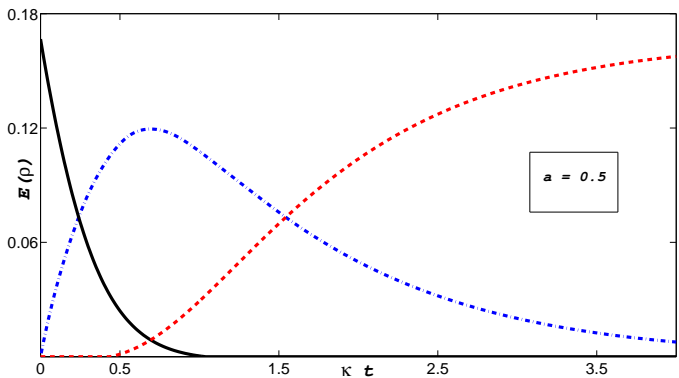}}
\caption{Same caption as Figure \ref{FIG:YEc1} but for $a = 0.5$.}
\label{FIG:YEc2}
\end{figure}

In Figure \ref{FIG:YEc2}, we show the entanglement monotones for the initial state (\ref{Eq:YES}) with $a = 0.5$. The cavity qubits become disentangled at 
$\kappa t \approx 1.04$, whereas the reservoir qubits get entangled at $\kappa t \approx 0.44$. The four qubits genuine entanglement again decays 
asymptotically after reaching its maximum value.

$(4)$ We consider mixed states
\begin{equation}
\rho_c = c \, |\phi\rangle\langle \phi| + (1-c) \, |11\rangle\langle11|\,, 
\label{Eq:rhoc}
\end{equation}
where $|\phi \rangle = (1/\sqrt{2}) \, (|00\rangle + | 11\rangle)$. These states bear some resemblance to states (\ref{Eq:JIS}) which exhibited the sudden change behavior. 
The time of ESD for these states is given by
\begin{eqnarray}
t_{ESD} (\rho_c) = \frac{1}{\kappa} \ln\bigg[\frac{2 - c}{2 (1-c)} \, \bigg]\, . 
\label{Eq:TESDa}
\end{eqnarray}
The corresponding time of ESB is given by
\begin{eqnarray}
t_{ESB} (\rho_c) = \frac{1}{\kappa} \ln\bigg[\frac{2 - c}{c} \, \bigg]\, . 
\label{Eq:TESBa}
\end{eqnarray}
These times are equal for $c = 2/3$ while, for $c < 2/3$, ESD comes earlier than ESB, and vice versa.

\begin{figure}[h]
\scalebox{2.20}{\includegraphics[width=1.95in]{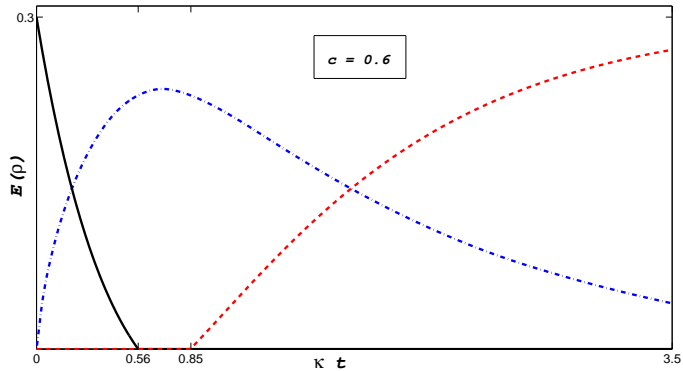}}
\caption{Entanglement monotone is plotted for cavity-cavity (solid line), reservoir-reservoir (dashed line), and four qubits  
(dashed-dottd line) for states (\ref{Eq:rhoc}) for $c = 0.6$.}
\label{FIG:Ex4c1}
\end{figure}
Figure \ref{FIG:Ex4c1} depicts entanglement monotone for cavity qubits (solid line), reservoir qubits (dashed line) and combined state of four qubits 
(dashed-dotted line) for a choice of $c = 0.6$. The cavities get disentangled at $\kappa t_{ESD} \approx 0.56$ whereas the reservoirs entanglement appear 
at $\kappa t_{ESB} \approx 0.847$. The combined state of four qubits shows no sudden change in its dynamics and decay asymptotically after reaching its maximum value. 

\begin{figure}[h]
\scalebox{2.20}{\includegraphics[width=1.95in]{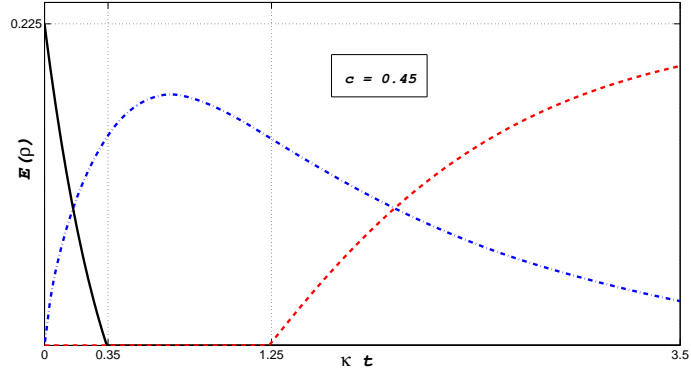}}
\caption{Same caption as Figure \ref{FIG:Ex4c1} but for $c = 0.45$.}
\label{FIG:Ex4c2}
\end{figure}

Figure \ref{FIG:Ex4c2} shows entanglement monotone for cavity qubits (solid line), reservoir qubits (dashed line) and combined state of four qubits 
(dashed-dotted line) for a choice of $c = 0.45$. The cavities become disentangled at $\kappa t_{ESD} \approx 0.35$, whereas entanglement among 
reservoirs appear at $\kappa t_{ESB} \approx 1.25$. We observe no sudden change in the dynamics of combined state of four qubits. Contrasting these last three figures, we observe that whether $t_{ESD}$ is larger or smaller than $t_{ESB}$, the four-qubit monotone behaves similarly and does not show a constant plateau value, rising only to a peak and then falling off monotonically.
 
$(5)$ Having observed the sudden change phenomenon only in one family of specific initial states, we investigated how much white noise 
can be tolerated by these states before the phenomenon is washed out. Consider the initial states 
\begin{equation}
\rho_f = f \, |\tilde{\psi}\rangle\langle \tilde{\psi}| + \frac{1-f}{4} \, I_4 \, ,
\label{Eq:SCWN}
\end{equation}
where $|\tilde{\psi}\rangle = (1/\sqrt{26}) \, |00\rangle + (5/\sqrt{26}) \, |11\rangle$ is the state which exhibits sudden change in dynamics of genuine entanglement 
as depicted in Figure \ref{FIG:CR4}. The states (Eq.\,\ref{Eq:SCWN}) are entangled in the range $13/23 < f \leq 1$. 
The time of ESD for these states is given by
\begin{eqnarray}
t_{ESD} (\rho_f) = \frac{1}{\kappa} \ln\bigg[\frac{13 + 37 \, f}{26 + 14 \, f} \, \bigg]\, . 
\label{Eq:TSDf}
\end{eqnarray}
The corresponding time of ESB is given by
\begin{eqnarray}
t_{ESB} (\rho_f) = \frac{1}{\kappa} \ln\bigg[\frac{13 + 37 \, f}{23 \, f - 13} \, \bigg]\, . 
\label{Eq:TSBf}
\end{eqnarray}

\begin{figure}[h]
\scalebox{2.20}{\includegraphics[width=1.95in]{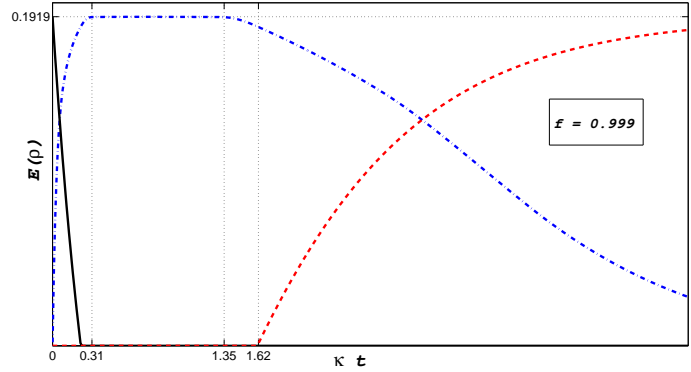}}
\caption{Entanglement monotone is plotted for cavity-cavity (solid line), reservoir-reservoir (dashed line), and four qubits  
(dashed-dottd line) for states (Eq.\,\ref{Eq:SCWN}) for $f = 0.999$.}
\label{FIG:Ex5c1}
\end{figure}
Figure \ref{FIG:Ex5c1} depicts entanglement monotone for cavity qubits, reservoir qubits and combined state of four qubits for $f = 0.999$. It seems that 
sudden change in the dynamics of genuine entanglement occurs for $0.31 \leq \kappa t \leq 1.35$, however a close examination of data reveals that it is not the 
case. The amount of initial entanglement for cavity qubits is $E(\rho(CC)) \approx 0.191865$, whereas the maximum amount of genuine entanglement achieved is  
$E(\rho(CCRR)) \approx 0.191783$ at $\kappa t \approx 0.69$. Actually in the range $0.33 \leq \kappa t \leq 1.32$, the amount of genuine entanglement varies only 
at 5th and 6th places after decimal point, that is, $E(\rho(CCRR)) \approx 0.1917xy$. Effectively, this change is too small to be noticed in Figure \ref{FIG:Ex5c1}. 
Hence, even a tiny mixture of white noise seems to wash out the sudden change phenomenon.

\section{Discussion and Summary} \label{Sec:conc}

We analyzed the dynamics of genuine multipartite entanglement in cavity-reservoir settings. We wanted to explore what happens to entanglement when it disappears between the two cavities, how it manifests in the reservoirs and in the full four-qubit system. Although some systematics are seen, especially that the four-qubit entanglement rises to a peak value and falls off asymptotically, there is no generic relation between the entanglement of the two cavities and the two reservoirs, appearance of it in the latter not simply related to its disappearance in the former. We found that for one specific class of 
initial states, the entanglement monotone for four qubits exhibited a sudden change in its dynamics. In a time window 
in which neither cavity-cavity qubits nor reservoir-reservoir qubits are entangled, genuine multipartite entanglement 
exhibited an interesting feature of an abrupt leveling off to a constant value during that time window before 
decaying. By enlarging the duration of this time window, we observed 
that the sudden change of multipartite entanglement also enlarged to the same duration. 
We then investigated other initial states but they do not exhibit this phenomenon. These observations indicate that this sudden change 
in the dynamics of multipartite entanglement is not a generic feature like sudden death or sudden birth of entanglement.
 
An interesting observation is the fact that the multipartite entanglement, 
although different from bipartite entanglement in a fundamental way, attains a maximum value either during the time window where neither 
cavity qubits not reservoir qubits are entangled, or at the point where the curves for cavity and reservoir entanglement intersect each other. 
Another interesting feature is that the multipartite entanglement always starts growing 
immediately after the interaction starts and reaches some maximum value. After this point it always decay asymptotically. Therefore, all 
four qubits remain genuinely entangled for a long time. We note 
that as the multipartite entanglement keeps decaying, the entanglement between reservoirs keeps growing until it attains the maximum value of entanglement available for this closed system of four qubits. In addition, the sudden change in dynamics of genuine entanglement is extremely sensitive to white noise. Even a tiny fraction of white noise washes out this effect. At $\kappa t = \infty$, the joint state factors  into $\rho(t = \infty) = |00\rangle_{C_1C_2}\langle 00| \otimes \rho_{R_1R_2}$ with all entanglement transferred to the reservoirs. 
\begin{acknowledgments}
M. Ali acknowledges discussions with Drs. Gernot Alber and Tobias Moroder. M. Ali is also thankful to Gernot Alber for his kind hospitality at 
TU Darmstadt where part of this work was done. M. Ali is grateful to Dr. Otfried G\"uhne for his generous hospitality at Universit\"at Siegen. 
This work has been supported by the EU (Marie Curie CIG 293993/ENFOQI) and the BMBF (Chist-Era Project QUASAR). A.R.P. Rau thanks Dr. J. C. Retamal for discussions and hospitality.
\end{acknowledgments}

\end{document}